\documentclass[acmsmall]{acmart}
\usepackage{tabularx}
\usepackage{array}
\usepackage{quotes}
\AtBeginDocument{%
  \providecommand\BibTeX{{%
    \normalfont B\kern-0.5em{\scshape i\kern-0.25em b}\kern-0.8em\TeX}}}

\AtBeginDocument{%
  \providecommand\BibTeX{{%
    Bib\TeX}}}

\setcopyright{acmlicensed}
\copyrightyear{2025}
\acmYear{2025}
\acmDOI{10.1145/3711051}

\begin{document}
\setcopyright{acmlicensed}
\acmJournal{PACMHCI}
\acmYear{2025} \acmVolume{9} \acmNumber{2} \acmArticle{CSCW153} \acmMonth{4}\acmDOI{10.1145/3711051}

\title[Pseudo-Automation]{Pseudo-Automation: How Labor-Offsetting Technologies Reconfigure Roles and Relationships in Frontline Retail Work}


\author[]{Pegah Moradi}
\email{pegah@infosci.cornell.edu}
\orcid{}
\affiliation{
\department{Department of Information Science}
  \institution{Cornell University}
  \city{Ithaca}
  \state{NY}
  \country{United States}}
  
\author[]{Karen Levy}
\email{karen.levy@cornell.edu}
\orcid{}
\affiliation{
\department{Department of Information Science}
  \institution{Cornell University}
  \city{Ithaca}
  \state{NY}
  \country{United States}}

\author[]{Cristobal Cheyre}
\email{cac555@cornell.edu}
\orcid{}
\affiliation{
\department{Department of Information Science}
  \institution{Cornell University}
  \city{Ithaca}
  \state{NY}
  \country{United States}}


\renewcommand{\shortauthors}{Pegah Moradi, Karen Levy, and Cristobal Cheyre}

\begin{abstract}
 Self-service machines are a form of pseudo-automation; rather than actually automate tasks, they offset them to unpaid customers. Typically implemented for customer convenience and to reduce labor costs, self-service is often criticized for worsening customer service and increasing loss and theft for retailers. Though millions of frontline service workers continue to interact with these technologies on a day-to-day basis, little is known about how these machines change the nature of frontline labor. Through interviews with current and former cashiers who work with self-checkout technologies, we investigate how technology that offsets labor from an employee to a customer can reconfigure frontline work. We find three changes to cashiering tasks as a result of self-checkout: (1) Working at self-checkout involved parallel demands from multiple customers, (2) self-checkout work was more problem-oriented (including monitoring and policing customers), and (3) traditional checkout began to become more demanding as easier transactions were filtered to self-checkout. As their interactions with customers became more focused on problem solving and rule enforcement, cashiers were often positioned as \textit{adversaries} to customers at self-checkout. To cope with this perceived adversarialism, cashiers engaged in a form of \textit{relational patchwork}, using techniques like scapegoating the self-checkout machine and providing excessive customer service in order to maintain positive customer interactions in the face of potential conflict. Our findings highlight how even under pseudo-automation, workers must engage in relational work to manage and mend negative human-to-human interactions so that machines can be properly implemented in context. We recommend future CSCW scholars consider the particular conditions that lead to relational patchworking demands on frontline workers. Understanding these conditions can help scholars and practitioners to more fully account for the multivalent effects of these technologies on workers, in retail and beyond.
\end{abstract}



\received{January 2024}
\received[revised]{July 2024}
\received[accepted]{October 2024}

\maketitle


\section{Introduction}
"`Unexpected item in the bagging area' is a shared cultural reference like no other,'' writes the journalist Kaitlin Tiffany, referencing a common self-checkout audio cue in order to encapsulate how mishaps and interruptions at self-checkout became a routine part of the American shopping experience  \cite{tiffany_wouldnt_2018}. The common shortcomings of self-checkout make it an example of what economists Acemoglu and Restrepo call ``so-so technology,'' or technology that lowers the demand for labor without any strong gains in consumer welfare or productivity, threatening long-term employment and wages  \cite{acemoglu_automation_2019}. While labor-offsetting systems like self-checkout or automated call centers are often purported to create efficiencies and reduce costs for firms, they often create other bottlenecks and inefficiencies that require human intervention  \cite{mull_self-checkout_2023, andrews_end_2018}. In this paper, we argue these technologies fail to reduce the reliance on human intervention, instead creating new forms of relational work that workers must engage in to manage the transformation of their roles and relationships.

We focus on self-checkout as an exemplary case of a widely deployed so-so technology in order to analyze how the nature and composition of frontline retail work has changed. In addition to economic work on the labor-market effects of these technologies, scholarship in CSCW and organization studies has long considered how workplace technologies reshape worker roles, tasks, and relationships  \cite{barley_technology_1986, orlikowski_using_2000, kellogg_algorithms_2019, suchman_making_1995}. More recently, the onset of the COVID-19 pandemic has led to a surge in investment in automating and labor-offsetting technologies \cite{chernoff_covid-19_2020, obrien_we_2021}. Still, little empirical work has considered how \textit{self-service} technologies in particular have changed the tasks and roles of the millions of frontline workers in the U.S. It is crucial to study the impacts of these popular workplace technologies: A deeper understanding can reveal the material effects of technological changes beyond sheer job loss and elucidate how digital technologies shape social interaction in collaborative work more broadly.

This study aims to contribute to existing research on technological change in frontline work by considering how offsetting labor from an employee to a customer via technology — or ``pseudo-automation'' — changes the composition and quality of work for retail workers. In this study, we investigate the following research question: \textbf{How do self-checkout technologies reconfigure frontline retail workers' roles and relationships to customers?} To address this question, we interviewed 22 current and former retail cashiers who had worked with self-checkout technologies. We find that self-checkout affected two dimensions of cashiering work tasks at self-checkout, and one at traditional checkout: When monitoring self-checkout, cashiers now dealt with \textit{parallel demands} from multiple customers at once and took on more \textit{problem-oriented tasks}, as their work involved intervening to assist customers and prevent thefts. Workers staffing traditional checkout lanes at stores that had both kinds of checkouts faced \textit{more strenuous transactions}, as customers with easier transactions were filtered into self-checkout. We find that the task changes at self-checkout in particular changed the \textit{relational} dimensions of cashiering work, such that cashiers took on a more adversarial role while monitoring self-checkout in order to enforce rules. To help manage these adversarial relationships in their customer-service roles, cashiers applied two main techniques, which we call \textit{scapegoating} and \textit{hyperservicing}.

Our findings contribute empirical insight into how systems intended to automate or offset frontline labor can create adversarial relationships between workers and customers, which workers must develop techniques to manage. We refer to this conflict-management work as \textit{relational patchworking} --- based off of Fox et al.'s notion of patchworking in human-AI workplace interactions  \cite{fox_patchwork_2023}. Relational patchworking demonstrates that pseudo-automation still requires skilled human labor to manage and mend interactions in frontline contexts. We suggest that CSCW scholars further consider these relational changes and the specific conditions and technologies that necessitate relational patchwork, allowing for worker- and customer-friendly technologies to be developed and deployed in the workplace.

\section{Background}
In order to address questions surrounding changes in tasks, roles, and relationships in retail work, we build upon previous scholarship in CSCW on invisibilized work in digital automation, bringing these works in conversation with related literature on retail and frontline work in economic and organizational sociology. 

\subsection{Self-Checkout and the Origins of Offsetting Retail Work}

Self-checkout is a decades-old workplace technology: Major retailers in the U.S. originally piloted self-checkout in the 1990s, with the machines growing in popularity in the 2000s  \cite{lake_how_2002, grimes_just_2004}. From 1999 to 2003, the percentage of supermarkets with self-checkout rose from 6\% to 38\%  \cite{grimes_just_2004}. Following the onset of the COVID-19 pandemic, self-checkout implementation expanded further as retailers responded to a decreased labor supply and public-health measures to limit in-person human interactions. By one industry measure, 96\% of grocery stores in the US now offer self-checkout lanes, comprising 29\% of all grocery transactions \cite{food_marketing_institute_food_2023}.

Self-checkout is part of a long history of retailers offsetting labor and control from employees onto customers. Grocery shopping in the 19th and early 20th centuries, for instance, typically involved a customer giving a clerk a list of items, which the clerk would measure and package for the customer in a labor-intensive process  \cite{andrews_-it-yourself_2009}. In 1916, Clarence Saunders opened the first Piggly Wiggly location, a grocery store concept that cut out clerks entirely and allowed customers to roam the store and pick pre-packaged items off of store shelves \cite{freeman_clarence_2011}. Advancements in individual product packaging, UPC barcodes, and electronic scanners led to customers having greater control in the shopping experience and made the Piggly Wiggly model of retail stores — one where customers take on most tasks involved in in-person shopping — commonplace in the U.S. \cite{andrews_-it-yourself_2009}.
\subsection{Automation and Human Labor in Contemporary Checkout}
Just as the Piggy Wiggly store concept offsets work from clerks onto customers, self-checkout offsets a cashier's tasks to an unpaid, untrained customer through a machine designed to make checkout tasks as intuitive as possible  \cite{ekbia_heteromation_2017,andrews_end_2018,litwin_understanding_2022}. In most cases, self-checkout machines employ a customer-facing touch-screen monitor, a scanning area, and a bagging area to allow the customer to go through the checkout process with minimal interference from store employees. The interface can limit what customers are able to do without employee permission, like removing a previously scanned or mis-scanned item, or purchasing age-restricted products like alcohol. While typical self-checkout machines have the customer scan product barcodes, newer self-checkout systems use other methods to detect which items are being purchased. Checkouts at the clothing store Uniqlo, for instance, use RFID tags to detect what items are being purchased, meaning customers no longer have to scan items in the checkout process \cite{lin_uniqlos_2023}. 

Some commentators consider self-checkout to be a stepping stone on the way to a fully autonomous retail store, where a customer doesn't have to interface with any frontline employee at all \cite{tiffany_wouldnt_2018, kingson_future_2022}. The most notable attempt at a fully autonomous checkout system has been Amazon's ``Just Walk Out'' product, which uses a series of cameras and sensors to track an individual throughout the store, detect what items they pick up or put down, and charge their account accordingly once they exit \cite{bitter_amazons_2024}. Even Just Walk Out, however, has been criticized for being covertly reliant on human workers: \textit{The Information} reported in 2024 that roughly 70\% of transactions in 2022 had to be manually reviewed by a team in India comprising roughly 1,000 human workers \cite{bitter_amazons_2024}. Amazon denied the reports, stating that Just Walk Out does not rely on ``human reviewers watching from afar,'' but rather that these human workers are doing standard AI labelling and annotation work \cite{kumar_update_2024}. (It's important to note, however, that even Amazon Go stores have frontline workers who stock items and—as one job posting for the role notes—``delight customers by answering questions and offering up recommendations as needed''\cite{amazon_amazon_2023}.)

\subsection{Challenges of Offsetting Work to Customers}
Customers are still untrained and less sanctionable than store employees, despite being assigned more and more retail tasks over time. At self-checkout, customers are thus more prone to mis-scanning or stealing items. To mitigate these issues, self-checkout machines often include technological loss-prevention systems, such as weight-checking (sensors that match the weight of the item whose barcode was scanned with that of the item placed in the bagging area) and public-facing monitors (screens that show the customer that they are being recorded) \cite{beck_self-checkout_2018, wells_pardon_2024, boudier_effective_2024, mortimer_watch_2020}. Other systems involve more cutting-edge technology: In some cases, self-checkouts use computer vision algorithms to analyze in real-time what items are being purchased, whether what is scanned matches what was bagged, and whether items are left behind in one's cart  \cite{nagaraj_computer_2023, proppos_proppos_2024}.

\begin{figure}
    \centering
    \includegraphics[width=0.5\linewidth]{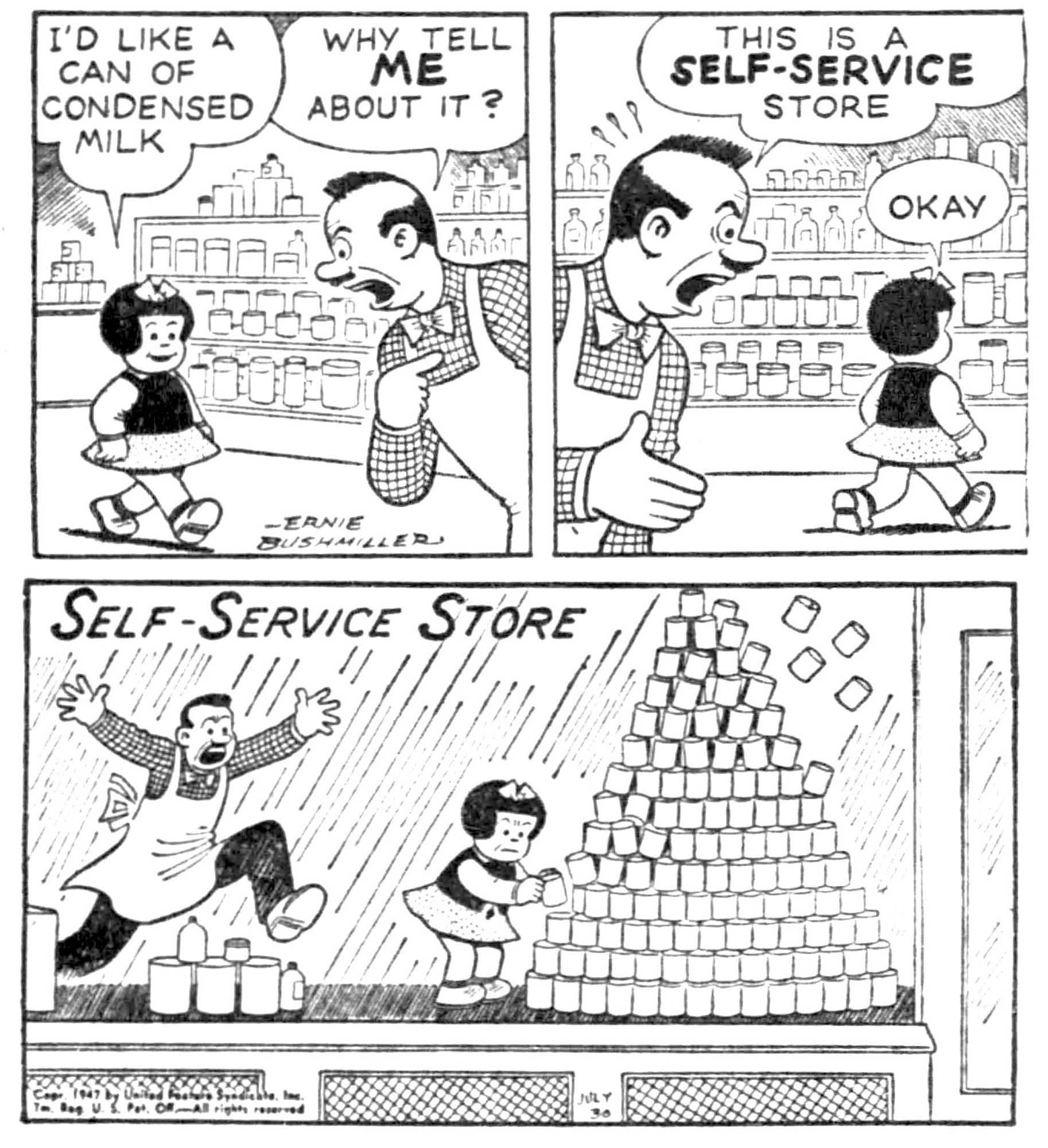}
    \caption{A 1947 comic from \textit{Nancy} by Ernie Bushmiller \cite{bushmiller_nancy_1947}. The comic depicts how offsetting work in a self-service store to untrained customers can lead to more accidents that workers need to rectify.}
    \label{fig:nancy}
\end{figure}

While these tools are specific to checkout, they are an extension of a much larger loss-prevention infrastructure that exists throughout most major retail stores. These tools, employed both to avoid loss and theft, as well as to collect analytics for marketing purposes and in turn monitor employee productivity \cite{levy_refractive_2018}, have also adapted as more and more work in the store is offset to customers. Stores employ a variety of anti-theft techniques ranging from locking up or placing antitheft tags on high-value, high-theft items \cite{dickinson_rfid_nodate, beck_self-checkout_2018}, to hiring uniformed or plainclothes asset-protection personnel, to receipt-checking, to implementing facial-recognition systems that identify and apprehend repeat shoplifters \cite{satariano_barred_2023}.  

In some cases, retailers have seemingly decided that the costs associated with self-checkout outweigh its supposed labor-saving benefits. Major retailers like Target, Walmart, and Dollar General have scaled back on the availability of self-checkout lanes in recent months, often citing losses and theft at self-checkout as justification for the change \cite{grothaus_walmart_2024, meyersohn_dollar_2024}. Self-checkout is thus a strategic context for investigating how labor-offsetting technologies change frontline work due to its unsettled role in modern retail, corporate hesitation with its implementation, and the tension retailers face between offsetting labor onto customers and simultaneously minimizing losses from customer errors and theft.

\subsection{Pseudo-Automation}
While the high-level promise of automation is to eliminate labor costs and improve efficiency, automation often involves new sources or forms of human labor required to create and maintain supposedly autonomous --- or pseudo-autonomous --- systems, like self-checkout. These kinds of dynamics are a specific case under a system of labor relations that Nardi and Ekbia refer to as ``heteromation'' \cite{ekbia_heteromation_2017}. Under heteromation, new technologies necessitate human labor to extract value, while also devaluing the very human labor needed to sustain the system \cite{ekbia_heteromation_2017}. HCI and CSCW scholars have long studied this kind of human labor in the form of ``invisible work,'' in which certain tasks and workers can be hidden from the production process and thus left unacknowledged and undercompensated \cite{star_layers_1999,suchman_making_1995, irani_turkopticon_2013, gray_ghost_2019}. The phenomenon of ``ghost work,'' as coined by Gray and Suri, extends this idea, elucidating that the maintenance of seemingly streamlined platforms relies on a vast network of low-wage human workers to label training data, handle edge cases, and fix errors in algorithmic outputs  \cite{gray_ghost_2019}.

Other scholarship considers how forms of invisible work help to integrate pseudo-automating technologies within their particular contexts. A 2019 Data \& Society report by Mateescu and Elish specifically considers grocery retail as a case study of how humans are necessary for the context-dependent integration of machines, writing that the introduction of customer labor in self-checkout has ``not reduced but rather reconfigured the skills and responsibilities that fall on cashiers'' \cite{mateescu_ai_2019}. Similarly, in their study of the implementation of AI technologies in frontline janitorial and waste-management work, Fox et al. introduce the concept of patchworking to describe the labor involved in making sure AI systems actually work as intended in their given context \cite{fox_patchwork_2023}. They identify three kinds of patchworking: \textit{compensating patchwork} (labor that makes up ``for the failures of machines, stepping in to perform their duties''), \textit{observation patchwork} (``babysitting the robots,'' or monitoring machines for moments of breakdown), and \textit{collaborative patchwork} (making sure the machine is aligned with the context in which it is deployed, often by tweaking the context to fit to the machine's needs) \cite{fox_patchwork_2023}.

Fox et al. explicitly note that the workers who engage in patchwork are often those whose work is ``closest to the forms of labor approximated by AI — workers whose tasks are offset when automation functions properly and whose roles are shifted when it fails''  \cite{fox_patchwork_2023}. The human labor involved in maintaining an autonomous system that is intended to replace those very workers is closely tied to Vertesi et al.'s notion of ``pre-automation:'' a process where firms hire outsourced labor forces without employment benefits (allowing the firm to offset business risks onto individual workers \cite{dubber_future_2020}) while simultaneously investing in technologies to replace them in the future  \cite{vertesi_pre-automation_2020}. These workers, typically working in platform or logistics industries like rideshare or delivery driving, often work under regimes of intense data collection — data which is later used to train their replacements  \cite{vertesi_pre-automation_2020}. 

Just as human labor is embedded throughout AI-driven automation — whether during, pre, or post-automation — it is also key in powering self-service systems, whether or not they include AI-driven technology. While retailers and industry analysts often describe self-service technologies like self-checkout as a form of ``automation''  \cite{begley_automation_2019}, most common self-service technologies don't really \textit{automate} worker tasks, but rather offset them to a customer, promising faster service and greater convenience  \cite{andrews_end_2018}. Pseudo-autonomous systems like self-checkout thus not only still require non-employee labor, they also need employees to stay in-the-loop to supervise and monitor the customer-users, similar to the contexts documented by Fox et al.  \cite{fox_patchwork_2023} and Gray and Suri  \cite{gray_ghost_2019}.

\subsection{Technological Change in Frontline Retail Work}
Frontline workers act as the primary interface between a bureaucratic organization and its clients or, in the case of retail, its customers  \cite{lipsky_street_1980}. Frontline workers execute the rules of bureaucratic organizations, often exercising discretion in how and when to administer these rules, making them an integral part of how regulations are practiced and enforced  \cite{lipsky_street_1980, maynardmoody_streetlevel_2010}. Despite their importance and outward visibility, frontline workers are historically ``rewarded the least, valued the least, and considered the most expendable and replaceable'' \cite[p. 176]{kanter_life_1979}. 

Frontline retail work in particular is often low-wage and made increasingly precarious as workers are subject to workplace data collection and automated management  \cite{van_oort_emotional_2019}. Their customer-facing position often exacerbates this, as data collected about customers can be used to analyze and manage workers, in what Levy and Barocas call ``refractive surveillance''  \cite{levy_refractive_2018}. In retail, for example, data collected to monitor customer behavior and provide marketing analytics can be repurposed to analyze worker productivity. Spektor et al. likewise review ways in which service work in the hospitality sector is increasingly mediated by digital technologies, including self-service, and thus enables greater algorithmic management of service workers  \cite{spektor_charting_2023}. Forms of algorithmic management like just-in-time scheduling can make service work overwhelming; workplace monitoring and algorithmic management necessitate another layer of awareness and anxiety — what Van Oort calls ``the emotional labor of surveillance'' — that exacerbates the difficulties of already-stressful service work \cite{van_oort_emotional_2019}. These difficulties and stressors can also be associated with self-checkout: As Eom and Schneider find in a survey of retail workers, workers at stores with self-checkout report greater customer incivility than workers at stores without it \cite{eom_please_2024}. In spite of these pressures, frontline retail workers still have to provide excellent customer service --- ``service with a smile'' --- and regulate their emotions at work in order to perform their roles \cite{grandey_emotional_2019, hochschild_managed_2003}.

Just as managers surveil frontline workers, we can expect customers to be increasingly surveilled as well, as more and more tasks are offset onto them and mediated by technological systems. But when customers act as workers, who watches over them? We draw on work in organization studies demonstrating how new technologies can restructure roles in other contexts, ranging from radiology departments  \cite{barley_technology_1986} to trucking \cite{levy_data_2023} to pharmacies \cite{barrett_reconfiguring_2012}. Self-service work can similarly reposition frontline workers to take on this role of customer-monitor and rule-enforcer as a result. Despite the recent wealth of research surrounding how new technologies can surveil, police, and punish workers  \cite{levy_data_2023, van_oort_emotional_2019, spektor_charting_2023}, little empirical work considers how self-service in particular is poised to reconfigure the roles of frontline retail workers. While job loss and algorithmic management are a part of the story of technological change in frontline work, we consider how work might change for the approximately 7 million frontline retail workers in the US (encompassing cashiers and retail sales workers) whose work is being immediately impacted by these new technologies  \cite{bureau_of_labor_statistics_cashiers_2022, bureau_of_labor_statistics_retail_2022}. 

\section{Methods}

To investigate how self-checkout affected frontline retail workers' roles and interactions, we conducted 22 interviews with current and former frontline retail workers across the United States. These procedures were approved by our institution's Institutional Review Board. 

\subsection{Sampling}
Interviews took place over two rounds: The first round of 10 interviews was conducted in January 2023. Participants for the first round (P1-P10) were either recruited via a worker-rights non-profit organization, which sent our recruitment message and screening survey to their existing network of retail workers, or through a message posted on a Subreddit ``for workers in the retail space to come together and support each other'' (/r/retailhell) with the permission of its moderators. Because the participants in our initial sample may have had a higher likelihood of holding frustrated, anti-self-checkout, or anti-management views, we supplemented this sample with a second round of 12 interviews (P11-22), recruited through a Facebook advertisement targeted at users who had listed cashiering and related retail roles as their occupation on the site. We did not notice meaningful differences in the data collected between the two rounds of interviews.

Participants were selected if they were 18 years of age or older and had worked with self-checkout as a frontline retail worker for at least a year continuously. All workers we interviewed were current or former cashiers who conducted cashiering tasks in their day-to-day roles regardless of their job title, including two participants whose job titles included the term ``manager'' (P2, P22). These two participants, however, did not take on key decision-making responsibilities; their roles still primarily involved cashiering tasks, and their responses were consistent with those of non-manager participants. All cashiers in our sample had experience working both at the self-checkout stations and at traditional checkouts in the same store.

Participants were based across 15 different U.S. states and worked at a variety of different retail stores, including grocery stores, big-box retailers, home-goods stores, and dollar stores. Because retail work tends to be low-wage (the May 2022 Bureau of Labor Statistics estimate for the median wage of cashiers in the U.S. was \$13.58/hour \cite{bureau_of_labor_statistics_cashiers_2022}) and paid hourly, the cost of spending time to schedule and conduct an interview for most frontline retail workers is significant; to help offset this cost, we offered participants who completed an interview a \$50 digital Visa gift card as compensation. Participant demographics are summarized in Table ~\ref{tab:participants}.

\subsection{Interviews and Analysis}



Interviews were semi-structured. As the organizational scholar Mette Vinther Larsen writes of conducting interviews with workers, ``[M]aybe the best thing a researcher can do to prepare for the conversation is to let go of the illusion of being in control of the conversation. Instead, the researcher could focus on engaging in a symmetrical co-authoring conversation with the organisational member by being sensitive to what emerges as the conversation unfolds'' \cite{madsen_using_2018}. In order to best explore the variety of backgrounds and store types in our sample, we began with an initial set of broad questions and probed further as participants discussed the circumstances of their specific workplaces. We asked participants broad questions about their demographic and career background, the stores they worked in, how they interact with self-checkout and customers, and how they feel about their daily work. Based on a review of the trade literature on self-checkout and prior fieldwork conducted at retail trade shows (where retailers and vendors frequently discussed theft as a major cost associated with self-checkout) we also asked participants how they engage with anti-theft systems and processes at work.

We interpreted this data using a grounded theoretical approach, generating themes and developing theory iteratively as we collected data \cite{glaser_discovery_2017}: We created an initial coding scheme inductively from our question list and the initial interviews, then revised the questions and codes as we continued conducting and analyzing interviews; for instance, after multiple participants brought up how they are tasked with preventing loss and theft at self-checkout, we began to ask participants about what motivates them to stop loss and theft at stores. We identified patterns and then themes that emerged from these codes by visually grouping similar codes and concepts together.

Interviews were conducted via Zoom by the first author and ran from 26 minutes to 55 minutes (not including the initial participant consent review, which typically took 5-10 minutes). All but two interviews were audio-recorded; two participants preferred not to have their interviews recorded, in which cases we took live notes during the interview that were included in the coding process. To preserve participant anonymity, transcriptions were pseudonymized, and all recordings were deleted after transcription.

\subsection{Limitations}

Notably, while our sample provided rich data, it has limitations. In order to capture a wide variety of perspectives from cashiers across different retail environments, we recruited and conducted interviews virtually. As a result, our sample is biased towards workers who spoke English and had access to an internet-enabled device. Only two of our participants (P7, P19) were not actively working as cashiers at the time of our interview; our sample thus underrepresents perspectives from workers who may have left their frontline retail jobs for a variety of reasons, including labor displacement, stress, disability, or promotion. Retail work in particular tends to have a high turnover rate, so while our sample is useful in elucidating workplace changes for today's cashiers, it does not encapsulate a large swath of the retail workforce that only spends brief periods of time in frontline retail work before moving on to other occupations. Lastly, the relationship between frontline employees and customers is two-sided, and in this study we only capture insights from retail employees. We focus on employees because their interpretation of the customer-worker relationship is ultimately what drives their own behavior change, and because they likely spend more time interfacing with customers at checkout than most customers spend interfacing with frontline retail workers. Yet customers' perspectives are also important to understanding broader changes in roles and relationships in pseudo-autonomous systems, and offer fertile ground for future study.

\begin{table}[htbp]
  \centering
  \caption{Study Participants}
   \label{tab:participants}
    \begin{tabular}{cp{7.5em}ccp{3.75em}cc}
    \toprule
    \textbf{Participant ID} & \textbf{Store Type} & \textbf{Years in Retail} & \textbf{Gender} & \textbf{Race} & \textbf{Age} & \textbf{State} \\
    \midrule
    P1    & Big Box  & 3.5   & M     & Asian & 35-44 & N/A \\
    P2    & Health Foods & 16    & M     & White & 25-34 & MO \\
    P3    & General Store & 4     & F     & Black & 25-34 & NY \\
    P4    & Dollar Store & 2     & M     & Black & 25-34 & GA \\
    P5    & Grocery & 27    & F     & White & 45-54 & NV \\
    P6    & Dollar Store & 3.5   & M     & Black & 25-34 & NY \\
    P7    & Grocery; Big Box  & 2.5   & NB    & Black & 35-44 & AR \\
    P8    & Grocery & 1     & M     & White & 18-24 & PA \\
    P9    & Grocery & 3     & M     & Black & 25-34 & GA \\
    P10   & Home Goods & 2     & NB    & Black & 25-34 & TX \\
    P11   & Grocery & 3     & M     & Asian & 25-34 & CA \\
    P12   & Big Box & 3.5   & F     & White & 45-54 & FL \\
    P13   & Grocery & 3.5   & M     & Asian & 25-34 & PA \\
    P14   & Big Box & 12    & F     & White & 55-64 & GA \\
    P15   & Grocery & 5     & F     & Black & 35-44 & OH \\
    P16   & Big Box & 10+   & F     & Black & 45-54 & GA \\
    P17   & Grocery; Big Box & 6     & M     & Asian & 35-44 & AZ \\ \vspace{2mm}
    P18   & Convenience Store; Gas Station & 4     & F     & Hispanic & 35-44 & PA \\ \vspace{2mm}
    P19   & Big Box & 4     & F     & White, Hispanic & 18-24 & MD \\
    P20   & Big Box & 13    & M     & White & 35-44 & CT \\
    P21   & Big Box & 5     & M     & Black & 25-34 & SC \\
    P22   & Convenience Store & 13+   & M     & White & 45-54 & VA \\
    \bottomrule
    \end{tabular}%
  \label{tab:addlabel}%
\end{table}%

\section{Findings}

\begin{figure}[h]
  \centering
  \fbox{\includegraphics[width=0.7\linewidth]{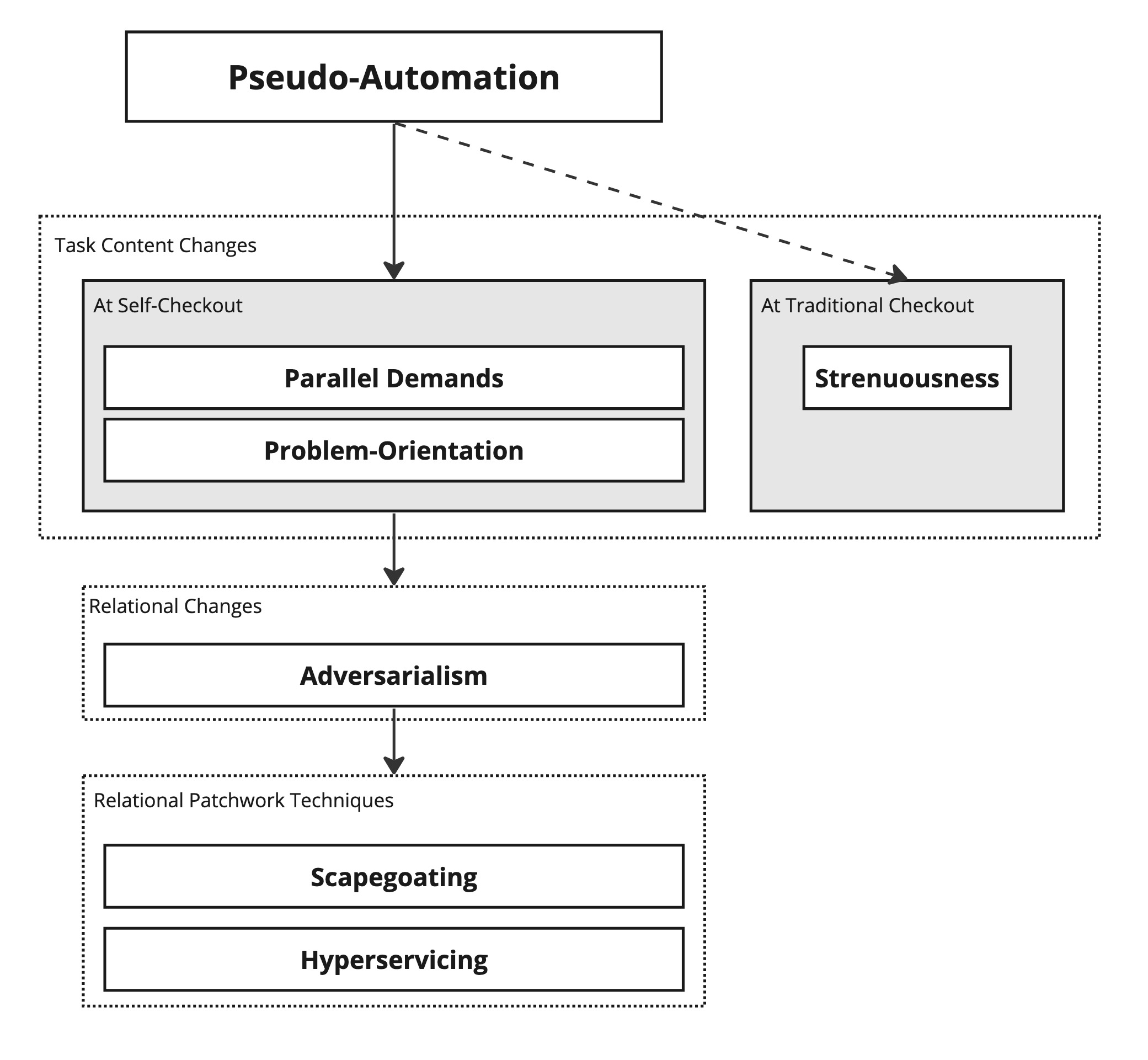}}
  \caption{Summary of findings}
  \label{fig:findings}
\end{figure}

While the workers we interviewed were aware of the possible risk of job loss, qualitative changes to their work were more top-of-mind. We find that self-checkout implementation changed two dimensions of tasks at self-checkout — which we describe as \textit{parallel demands} and \textit{problem-orientation} — as well as the increased strenuousness of traditional checkout. At self-checkout, these task changes led to a relational repositioning of cashiers as adversaries to customers. Cashiers then employed two techniques for handling adversarial customer interactions: \textit{scapegoating} the self-checkout machine for its flaws, as well as \textit{hyperservicing}, or masking surveillant behaviors under the guise of providing extra customer service. We summarize these findings in Figure \ref{fig:findings}.

\subsection{Preamble: Job Loss Concerns}

As an initial matter, we acknowledge the oft-discussed issue of job loss due to self-checkout. We found that though the cashiers we interviewed were well aware of the possible threat of job loss, they tended to keep these concerns in the periphery.

Cashiers often noted that they, their coworkers, or their customers had concerns about self-checkouts taking away cashiering jobs, typically bringing up these concerns before we asked about them explicitly. In one extreme example, a key manager at a health-foods store chain said he had heard of cashiers at another store location staging a walk-out in protest of that store's self-checkout implementation, citing concerns that they would be moved to other parts of the store or else would have their working hours ``\textit{drastically cut}'' [P2]. In some cases, participants said that management explicitly communicated that self-checkouts were added to prevent hiring new employees. As the store manager of a small convenience store on a military base noted: ``\textit{I have asked for an employee, my third employee, because two years ago we had three employees and she retired. And I was told that we would be able to get one, and then COVID of course happened, the sales went down and then they decided to add that [the self-checkout machine]. And the joke was, `Here's your third employee.' And that came from the general manager. So, it seems to be my third employee. I don't think I'll get my third employee back}'' [P22]. Cashiers were cognizant of possible employment declines in the long-term, both on store-wide and occupation-wide levels.

The vast majority of participants noted that job loss anxieties didn't affect how they approached career planning, with the exception of one participant, who started to take on more shifts doing stockroom and inventory work in order to pivot to what he believed would be higher-paying jobs in supply-chain logistics [P13]. Concerns were often assuaged either because the cashiers planned to only stay in their jobs in the short term [P11], or because they saw that they were still busy at work: For instance, a cashier at a gas-station convenience store that also served food remarked that the new self-checkout kiosks at her store freed her up to focus on other tasks that were part of her daily work, like making drinks and stocking items. Even with the self-checkouts, she noted, ``\textit{Overall, they're still going to need people…the store is busy enough}'' [P18]. Participants often cited feelings of understaffing and overwork prior to the implementation of self-checkout as a reason for dampened worries. As the key manager at the health-foods store explained, ``\textit{No one's hours got cut or anything in the store because of self-checkouts. The budget got slashed on the front-end but we were never using that budget, because we never had the people to use it, so no one really felt the effects of it taking away anything from anyone at the store}'' [P2]. Crucially, as discussed previously, because our sample only includes participants that kept their jobs after self-checkout was implemented at their store, we do not capture perspectives from workers who may have left their roles, either voluntarily or involuntarily.

\subsection{Effects on the Nature of Work Tasks}

While participants were aware of job loss as a potential threat, they were instead more focused on changes to their day-to-day work. We found that self-checkout systems spurred two changes for self-checkout attendants: First, cashiers now had to deal with more problems at once, responding to customers in parallel. Second, cashiers' work became increasingly problem-oriented, as opposed to the more routine, process-oriented work at traditional checkout. Work tasks at traditional checkout also changed, becoming more strenuous as customers took more difficult transactions through staffed checkout lanes.

\subsubsection{Parallel Demands at Self-Checkout}

Participants noted that a major challenge with self-checkout was the increased ratio of customers to employees in the checkout process: Whereas as cashiers they had one-to-one interactions with customers, as self-checkout attendants they had to observe and intervene on multiple checkouts—and therefore multiple customers—in parallel. One big-box store cashier described how this one-to-many relationship made work at self-checkout more hectic compared to traditional checkout: ``\textit{There's so many customers [at self-checkout] and I can only be with one customer at a time. So it's frustrating for customers and for myself…You feel pressured because they [management] want people moving through smoothly, and quickly. But if it's just me by myself it's a lot, because if I'm having to scan an ID or take discounts off, running back and forth to different registers, it's a lot. If there was another person, like another employee there with me, I wouldn't feel so much pressure and I wouldn't feel so frustrated}'' [P17].

Other cashiers described these parallel demands as making it more difficult to help prevent loss and theft at the store, now an important part of their monitoring role. The change in ratio decreases oversight, which makes theft easier, and surveillance more challenging, even in situations where the customers may not be planning on stealing items at self-checkout. As a former cashier and self-checkout attendant (who had since moved to the deli department) put it, ``\textit{I feel like versus if it is one cashier on the front side and one [customer] on the back side, I am saying you have more eyes on you and a person knows, like, okay, they are really watching. But, hell, if you've got one person watching six or seven people at one time, I am just saying, it is easier possibly for a person to say, I am not going to scan all these Cokes, I am not going to scan this dog food}'' [P7]. In some cases, customers actively exploit the limitations of this parallel task structure at self-checkout in order to steal: ``\textit{I feel like the ones that are going to steal,}'' a bookkeeper and cashier at a major grocery store chain noted, ``\textit{they are going to go through self-checkout because they know we can't watch four stations at one time […] They set up where they have somebody at another register and then they will purposely trip it off [trigger the self-checkout in a way that alerts the attendant] so that myself or another coworker has to go over and attend them}'' [P5].  Cameras and other antitheft devices are intended to help deter and catch theft, but these systems are often easily circumvented by savvy customers, and the employee is left to fill in the system's gaps.

Still, the parallelism at self-checkout did not always lead to constant chaos. Instead, the busyness of self-checkout was dependent on customer behavior. More customers checking out at a given time meant the attendant might suddenly be inundated with requests, while during lulls or slow periods the attendant could be standing by idly for long swaths of time. As one grocery-store cashier — who worked in a rural area of Pennsylvania with an older customer base — described, ``\textit{Most of the time, it [monitoring self-checkout] is boring just because nobody wants to use them. And then, when they do get busy, it is a little bit stressful because you have the machine going off constantly trying to fix everybody's transaction}'' [P8]. In some cases, the self-checkout attendants have to find ways to pass the time, ranging from organizing nearby merchandise displays [P8] to playing games on their phone—so long as their manager isn't around [P6].

Participants differed in whether they preferred the dynamism of self-checkout or the consistency of traditional checkout. The key manager at the midwestern health-foods store noted that most of his coworkers preferred staffing traditional checkout because it involved less idle time: ``\textit{I think most of them prefer to work at the normal register, just because you are always doing something at the normal register; whereas, the self-checkout, there is a lot of downtime where you are kind of bored. I think there are a few people that like that, but I think, for the most part, people like to be constantly engaged}'' [P2]. But others, like a former cashier at a big-box home-improvement store, found self-checkout to be more engaging, recounting that monitoring self-checkout ``\textit{was definitely more interesting, and because since they were four registers, when I got busy it was stressful, but it was kind of like, I felt like I wasn't just standing around doing nothing as much}'' [P19]. In part, this cashier noted that she found the dynamic pacing added variety and breaks to her work: ``\textit{I feel like the ebb and flow was nice because after the big rush, then it was like, you get the little break of `Okay, now I can start cleaning.' If anything spilled and stuff, wipe stuff down. So it was nice because it was after all the people come in, then you could start to take some time to check the registers for receipt tape and stuff}'' [P19]. Others—like P1, P8, and P3 — still preferred the comparative ease of ``\textit{standing there}'' [P1] and only having to intervene when there was an issue.

\subsubsection{Problem-Oriented Work at Self-Checkout}

 Cashiers noted that interactions with customers at self-checkout almost always began with a problem or a request for help: Monitoring self-checkout was focused on assisting customers, resolving errors, and preventing or intervening on thefts. A cashier at a dollar store in Georgia noted this difference: ``\textit{The cashier for me is basically, you know, easy to do: Talk to the customer, know what they want to buy, get the products from them [...], put it into the system, check the price, collect the money, and give them their change and then wish them a great day. But, you know, with the self-checkout system, sometimes the customers, they make mistakes with the system and it gives off errors and sometimes it is just really difficult to fix}'' [P4]. The problem-oriented tasks at self-checkout thus differed from the methodical, procedural task content of work in traditional cashiering. Even loss prevention at traditional checkout involved set procedures that cashiers were trained on, as the manager of the military convenience store explained: ``\textit{BOB and LISA: the `bottom of the basket,' `look inside for all items.'  Those are things that you as a retail worker are trained for, and your speed is quicker, even though you're scanning every item}'' [P22].

 Cashiers commonly noted that errors at self-checkout were more frequent, in part because customers were an unpaid, untrained source of labor. ``\textit{When customers are doing it, they're not being paid to do it. They don't feel the need that they have to do things correctly,}'' a cashier at a grocery store in southern California described. ``\textit{Rather than, when I do it, I'm on the clock, I'm being paid for my work. So, I definitely do feel the need to be correct}'' [P11]. At times, though the self-checkout machines were meant to be user-friendly, their technical limitations caused even more problems. False positives in anti-theft sensors, for instance, could pause transactions and require the cashier to step in before the customer could continue checking out. ``\textit{[The self-checkout machines] all go off of weight,}'' the cashier in rural Pennsylvania described, ``\textit{So, if the customer just moves something around, the entire thing is screwed up}'' [P8]. Other straightforward technical limitations — like age verification when purchasing alcohol or removing a previously scanned item if a customer changes their mind — still require the intervention of an employee to resolve the issue, adding to the set of possible problems that can arise at self-checkout.

 \subsubsection{Strenuous Tasks at Traditional Checkout}

 In addition to task changes that came from monitoring self-checkout, some cashiers described how working at traditional checkout became more challenging when both options were available for customers. Customers tended to self-sort themselves into different checkout options depending on what they were purchasing, taking easier, lower-quantity transactions to self-checkout. Customers thus tended to take their higher-quantity or complex transactions to traditional checkout. A designated self-checkout monitor at a big-box retailer in Georgia illustrated the difference in pressure between the two checkout lanes:
\begin{quote}
    \textit{\textbf{Q: } Is the flow of customers [at traditional checkout] pretty similar to self-checkout?\\
\textbf{A: } Oh, you get a lot more people for self-checkout because we have way less cashiers. So we get way more people in U-scan [self-checkout] any day than we would on the actual register, but the thing behind that is the people that we do get, they have a large, large sum of groceries.  Like we're talking 400. So you just ringing up and bagging a lot of stuff versus when you scan, I'm watching you bagging your own stuff.} [P16]
\end{quote}
Lifting, scanning, and bagging a large quantity of items in a single transaction could thus be more physically tiring. The increased physical demands of traditional checkout and the offsetting of responsibility made monitoring self-checkout easier and in many cases a more preferable role, but one that came with a downside for those on traditional checkout: While traditional checkout work remained more routine and procedural compared to the problem-oriented work of self-checkout, it also became more demanding for the cashier.

This filtering exacerbated existing pressures of staffing traditional checkout. Participants noted that there was greater pressure to move quickly at traditional checkout, in part because the cashier was the ``\textit{center of attention}'' [P12] in the interaction. One cashier explained how staffing traditional checkout was ``\textit{a little frustrating because it's like you don't have the help of somebody. I'm scanning it and bagging it.  There's nobody there to help you with that process, to help you bag all of that stuff.  So they [management] want a good flow where you're not taking too much time, but it's like, how do I do that?  So I would feel anxious and frustrated}'' [P17]. Big-box store workers explained that pressure came from both customers and management, as management would often keep track of how long transactions took at traditional checkout, and customers in line would get frustrated if they weren't checking others out quickly enough. ``\textit{I feel like it's more pressure working as the cashier},'' one big-box cashier explained, ``\textit{because we get timed too, and then people don't have patience — the next customer behind the customer I'm waiting on. In the self-checkout, sometimes it just flows a little faster}'' [P12]. Self-checkout, in comparison, offloaded some of the responsibility onto customers, and in turn relieved some of the pressure of worker productivity tracking at checkout.

\subsection{Resulting Relational Changes: Taking on an Adversarial Role}

At self-checkout, cashiers often felt positioned as adversaries to their customers, as their role had a greater focus on monitoring and rule enforcement than at traditional checkout. Even when they were simply assisting customers at self-checkout — not intervening on a potential theft — cashiers were usually coming to the customer at a tense moment, after a problem had already occurred. Some participants said they faced immediate hostility from customers when they came over to help with the self-checkouts as a result. As one grocery-store cashier recounted, ``\textit{when they make a mistake with the machine, maybe they did something wrong and then the machine is not responding and you know, they just kind of like get, I would say, pissed off. And some, they come ask me for help, while some just try to get angry and say the machine is a waste of time or something like that. And I am just there to try and make them calm and try to resolve their issue}'' [P9]. Self-checkout attendants thus take on an adversarial role of having to monitor and enforce rules, while also having to maintain positive customer interactions—both to protect the store's brand and customer base, as well as to protect themselves from potential verbal or physical abuse from customers.

Participants tended to be acutely aware of the risks that came from intervening on potential thefts, often having experienced abuse from customers in self-checkout first-hand. One cashier at a major grocery store chain described two instances that had occurred earlier that same day:
\begin{quote}
    \textit{So, a customer called me over to their station because they rang up the wrong lemons. They rang up organic lemons and not regular lemons. And then, when I looked over, I saw they already put two items into their bag. And I said, `Oh, did you want to pay for the two items you put in your bag?' And they proceeded, like, started cussing at me. And I was like, `Okay, I am just trying to help you with your lemons that you asked me to remove.' And then, they tried to pay for their order and then obviously the money wasn't there. So, they were highly upset with me. So, that was one order. And then, another order was two women that came through… Because we have to watch four of them at one time… And they left all their meat on the bottom of their cart. So, when they were scanning their items, they left the meat on the bottom, like, underneath their bags. So, I went over and I just asked them, I said, `Hey, did you guys want to pay for these items?' And they are, like, `Why are you watching us?' Like, very verbally abusive? Very verbally abusive.} [P5]
\end{quote}
In addition to these adversarial situations involving verbal abuse or conflict, employees also avoid accusing customers of theft outright out of fear of physical retaliation. Most participants said they received little training on how to prevent theft in the store, but learned over time how to respond to suspected thefts. Stores tended to advise cashiers not to intervene (purportedly to avoid liability and protect employees), yet cashiers remained the store's first line of defense against suspected thefts:
\begin{quote}
    \textit{\textbf{Q:}  Do they tell you to say anything to the customer or intervene in any way?\\
\textbf{A: } No.\\
\textbf{Q:}  Do they tell you not to do that or do they just not tell you…\\
\textbf{A: } Yes, most of the time, you are advised to just report it immediately.\\
\textbf{Q: } And why do you think they don't want you to intervene?\\
\textbf{A: } Because, at times, you cannot know if the customer is armed or not.} [P3]
\end{quote}

Even when situations didn't escalate to the point of physical or verbal abuse, monitoring customers sometimes made workers feel awkward or uncomfortable. Take, for instance, this response from a current self-checkout attendant at a dollar store in New York: ``\textit{[Monitoring self-checkout] feels stalky because you just have to watch and watch until they leave the self-checkout points. So, sometimes, even I myself feel cringey}'' [P6]. Having to take on an enforcement role felt out-of-character for some participants who were used to having a less confrontational role as a cashier, as explained by a former grocery store cashier who later chose to transfer to the deli department: ``\textit{I feel like, in some ways, you kind of already have it in your head that potentially someone is going to steal something or just not scan everything. And, me, I don't like being that type of person, I guess. I don't know… It is kind of weird, in a sense}'' [P7]. In order to maintain positive interactions, then, employees had to manage customer emotions as well as their own uneasy feelings about watching customers.

\subsection{Managing Adversarialism: Scapegoating and Hyperservicing}
In order to navigate having both an adversarial and supportive relationship to customers at self-checkout, participants described two primary methods of managing these tricky and potentially unsafe interactions. First, some participants described \textit{scapegoating} the machine as a way to defuse tension and avoid accusing the customer of theft. When the machine detected an instance of potential theft, attendants would come to check the situation and blame the machine for the alert. For instance, the now-deli worker remarked: ``\textit{I kind of say stuff like, `Well, you know how computers are,' or `Technology these days are not as'… Which, you know, they are not. You know, little stuff, something like that to kind of ease the time or that moment, I should say}'' [P7]. The store manager at the health-foods store described a situation where a customer left without paying for their items: ``\textit{Sometimes when we have noticed that a screen wasn't paid for and we notice that they are still walking out, we will run out and go grab them… `Hey, your card didn't go through or whatever,' and they will come back in and pay. I don't know if they were necessarily accidentally, or it was on purpose but that is just how we usually handle it.}'' [P2] Employees thus point out the technical limitations of the self-checkout machines as if to say it was not the customer who triggered the alert, but the technical system, relieving the customer of feelings of blame or accusation. This process can also put the customer and employee on the same side, against the machine, as evidenced by this explanation from the New York dollar-store cashier of how he approached customers facing issues with the self-checkout machine:
\begin{quote}
    \textit{\textbf{A: } I will try to explain to you that everything is automated and I am just like a help book or a manual. I am not the user. That everything is automated and it is the computers.\\
\textbf{Q: } And does that usually help the customer after you say that? How do you think they—\\
\textbf{A: } Most of the times when you say that, the next thing that comes out of their mouth is `Stupid computers.' }[P6]
\end{quote}

The second technique that cashiers used to defuse tension in high-stakes interactions at self-checkout was to perform surveillance as an extension of customer service, in what one worker at a home-improvement store referred to as ``\textit{customer servicing them to death}'':
\begin{quote}
   \textit{ We kind of fall back on just really customer servicing them to death. Asking how we can help them, if they're finding everything they need, what are you building with all your tools? Things like that, and trying to kill them with kindness, I guess. And a lot of times it works. A lot of times it works, if everybody's doing that to them, they kind of get the hint. We've recovered quite a few carts of pretty pricey stuff that they'll leave in the front end and just walk away and you kind of assume that they were planning on walking out the front door with stuff and they just leave it behind.} [P20]

\end{quote}

This kind of \textit{hyperservicing} is, of course, not unique to self-checkout. The cashier in rural Pennsylvania remarked that this technique (which he called ``\textit{aggressive hospitality}'') was applicable both at self-checkout and throughout the rest of the store: ``\textit{Like, say I noticed they are just taking something off the shelf and putting it into their purse or something,''} he explained, \textit{``Just, `Hey, can I help you with that? Can I help you find anything?' almost to the point where it is annoying}'' [P8]. While hyperservicing was useful for preventing loss and theft throughout the store, it was a crucial tool for cashiers to mask their monitoring and rule-enforcement at self-checkout as well.

\section{Discussion}

In this study, we find that offsetting work off of cashiers onto unpaid, untrained customers under pseudo-automation changes the task composition of work for remaining employees, making their tasks more problem-oriented and demanding in parallel, ultimately positioning workers as adversaries to customers, correcting their mistakes and enforcing anti-theft rules. In spite of their new adversarial role, employees still had to provide positive customer service experiences; they thus engage in techniques like hyperservicing and scapegoating in order to walk the line between these discrepant roles.

Our findings extend scholarship in CSCW and related fields in two key ways: First, we expand the realm of patchworking and configuration work beyond autonomous or AI systems to include pseudo-autonomous ones, where work is not actually automated but rather offset to an unpaid (often untrained) human. Second, we demonstrate that not only do pseudo-autonomous technologies change the task composition of work, but that by offsetting work tasks, the relational characteristics of frontline work can change as well. Understanding these changes enables scholars and practitioners to more fully account for the multivalent effects of these technologies on workers, even beyond the realm of retail.

In what follows, we develop the concept of relational patchworking to describe how frontline workers repair conflicts or emotional challenges that can arise when labor is offset to a human. We argue that without accounting for relational changes, our understanding of automation and pseudo-automation is incomplete. We then discuss how relational patchworking can help CSCW scholars better conceptualize interrelated systems of surveillance and other material conditions of different forms of frontline work.

\subsection{Relational Patchwork}

We use the phrase relational patchwork to describe the work involved in mending adversarial interactions brought on through gaps or failures in pseudo-automation, drawing on Fox et al.'s use of the term ``patchwork'' to describe the work involved in implementing AI systems in context.

In the three forms of patchwork Fox et al. describe (compensating, observation, and collaborative), workers integrate technologies into context through human-machine or human-object interactions. Workers, for example, can mop up water that an automatic floor cleaner leaves behind, or adjust the position of a bottle so an AI-driven recycling robot can properly pick it up. But under pseudo-automation, integrating technological systems in context also requires training, managing, and supervising \textit{other people}. Relational patchwork encapsulates the delicate and at times risky work of supervision, enforcement, and correction tasks that frontline workers must increasingly conduct in place of the work tasks that were offset to other people.

Patchworking as a framework intends to disrupt myths surrounding the ability of AI tools to enhance workplace productivity and deskill human work \cite{fox_patchwork_2023}. Pseudo-automation involves similar myths; as we demonstrate, the use of relational patchworking suggests that implementing pseudo-automation (whether or not it incorporates AI) still necessitates skilled human relational labor in frontline contexts, as offsetting labor creates social and emotional challenges that only human workers can properly assuage.

\subsection{Relational Patchworking Techniques and Worker Discretion}

How, then, did workers assuage these interpersonal challenges? The patchworking techniques cashiers described in our findings — hyperservicing and scapegoating — are not exhaustive, and frontline workers may engage in a variety of different strategies to attempt to maintain positive interactions with their clients. But the differences our respondents described between the two techniques reveal how technologically mediated interactions can change how frontline workers exercise discretion: Both are attempts to avoid accusing customers of misbehavior outright, but only hyperservicing involves the worker deciding whether or not they should step in to interrupt a customer and prevent possible theft. Workers engage in positive, conciliatory customer service behaviors to avoid appearing accusatory or threatening — tactics that can be applied throughout the store, not just at self-checkout.

Scapegoating, on the other hand, typically involves an interruption brought about by the machine itself — for instance an ``unintended item in the bagging area.'' Because self-checkout errs with some noticeable frequency (whether by incorrectly detecting certain items or barcodes, pausing transactions when it detects extra weight in the bagging area, or simply freezing as computers sometimes do) workers reacting to these interruptions leverage these problems even when the system might be working as intended. The employee can use the ambiguity and opacity of the system's alerts to avoid having to accuse the customer of theft or incompetence, ultimately giving the customer the benefit of the doubt.

Sociological theory dating back to Goffman has long explored how individuals employ these sorts of face-saving techniques in order to maintain their dignity and avoid conflict \cite{goffman_cooling_1952, goffman_face-work_1955}. Technological intermediaries can be fertile targets for blame, even outside of pseudo-autonomous systems: Hohenstein and Jung call this phenomenon ``AI as a moral crumple zone,''  flipping the term Elish originally used to describe how human operators absorb the blame for technological failures in high-stakes situations --- in the same way the crumple zone of a vehicle absorbs the impact of a collision \cite{hohenstein_ai_2020, elish_moral_2019}. Hohenstein and Jung instead find that in dyadic conversations, when participants had access to AI-suggested ``smart replies,'' they tended to attribute less responsibility to their conversational partner when conversations failed, suggesting that the AI absorbs some of the responsibility for the failure, regardless of whether or not it is actually at fault \cite{hohenstein_ai_2020}. The same kind of absorption of responsibility exists in workers' interactions with customers, mediated by the self-service machine, regardless of whether the machine actually erred.

Pseudo-automation in retail can therefore create some situations where workers have to exercise more discretion in proactively preventing potential theft (hyperservicing, when the technology may not catch errors and the worker must decide whether or not theft might be occurring), and other situations in which they react to technologically-mediated cues, conciliating and redirecting blame away from the customer (scapegoating, when the technology has supposedly caught an error and absorbs responsibility for the intervention).

\subsection{Relational Patchworking in the Future of Retail}

If the need for relational patchworking arises from the fact that machines err or offset work to untrained customers, what then happens if the self-checkout improves, whether by becoming more user-friendly, having the machine actually automate away checkout tasks (as Amazon Go or Uniqlo self-checkout are attempting), or by improving the accuracy of its loss- and theft-detection tools? In other words, what might happen if pseudo-automation can eliminate labor more effectively than it currently does? Though workers today often exploit the limitations of self-service machines to avoid conflict, retail technology vendors are increasingly developing and marketing new tools to improve self-checkout's ability to detect loss and handle edge cases, like initiatives to employ AI-based age verification, or to use RFID tags on all merchandise within a store in order to automatically detect what items are being taken in and out of the checkout area \cite{ncr_ncr_2018, dickinson_rfid_nodate}. 

While further work will be needed to assess how newer systems will work in practice, our findings suggest that such technological changes that lower the frequency of problem-oriented interactions and alleviate the parallel demands on workers could also alleviate the adversarial nature of frontline retail work by simply offsetting discretion onto a machine. However, if these systems become more and more accurate, it may also lead to even \textit{more} adversarial interactions, as technical systems may become a source of ``truth'' when identifying possible theft situations. Workers may thus lose the ability to offset some blame and pressure of enforcement by pinpointing the failures of the technology. That is, workers may no longer be able to rely on the black-box nature of these systems if they're broadly perceived by workers and customers to be extremely accurate. For these workers, transparency and improvements in the accuracy of the technology do not always alleviate interpersonal conflict; opacity and errors can instead be useful tools for cooling tense encounters.

\subsection{Bridging Patchwork and Surveillance}

Brick-and-mortar retail is rife with the tracking and monitoring of both workers and customers, whose surveillance is often intertwined (what Levy and Barocas describe as ``refractive surveillance'' \cite{levy_refractive_2018}) and technologically or, at times, algorithmically heightened (as in Van Oort's ``emotional labor of surveillance'' \cite{van_oort_emotional_2019}). Our findings suggest that when checkout tasks are offset to customers, workers take on more enforcement tasks. Frontline retail workers therefore are both surveilled by management while also surveilling customers.

Surveilling customers, as a new part of workers' roles, is more emotionally and relationally fraught than scanning items or conducting discrete tasks. Instead, workers and customers are situated in an interrelated web of surveillance. And just as emotional labor is involved in the process of being surveilled \cite{van_oort_emotional_2019}, surveilling customers also involves overlooked forms of relational work that are critical to ensuring these systems run as intended — that is, that customers don't create too much loss for the retailer. Relational patchwork thus draws our scrutiny to the interrelation between the surveillance of workers and customers (brought upon by new workplace technologies) and the effort required to manage these surveillant relationships.

\subsection{Future Work and Relational Patchwork in Other Frontline Contexts}

Of particular interest to CSCW scholars is the human labor that makes computer-supported collaboration possible, most notably in discussions of articulation work (``work that gets things back `on track'" \cite[p. 275]{star_sociology_1991}), invisible work, and ghost work \cite{gray_ghost_2019}, with the goal of better understanding ``how the [technology] affects relations of power and the nature of work'' \cite[p. 24]{star_layers_1999}. As discussed in section 2, scholarship like that of Fox et al. and Mateescu and Elish built upon this research in recent years by bringing to light the often overlooked configuration work involved in implementing AI systems within their intended contexts \cite{fox_patchwork_2023, mateescu_ai_2019}.

Our findings indicate that in retail, even pseudo-autonomous systems require relational patchworking in order to be implemented in context. This provokes one question of particular interest for future study: \textit{What workplace conditions and technological changes occasion these kinds of relational patchworking demands?}

Just as we have found in retail, one might imagine that similar dynamics could emerge in other workplaces, where frontline workers have multivalent, technologically-mediated relationships with customers or clients --- particularly those which involve both service and policing. In her study of frontline pharmacy workers, for instance, Chiarello explains how the ``discrepant institutional logics'' (conflicting logics of treatment in healthcare and punishment in law enforcement) brought about by the ``War on Drugs'' turn pharmacists into drug gatekeepers \cite{chiarello_war_2015}. Pharmacists thus take on an adversarial role against the patients they are also treating, a dynamic that Chiarello suggests can be exacerbated by national databases of controlled substance dispensation \cite{chiarello_war_2015}. Pharmacists then use avoidance techniques (saying a drug is out of stock or refusing to stock certain drugs) in order to skirt difficult scenarios in which they must decide whether to provide treatment or punish a patient \cite{chiarello_war_2015}. Similarly, Levy's work on electronic logging devices (ELDs) in trucking describes how law enforcement officers and truckers exploited the unfamiliarity or ``black box'' of the ELD to mediate previously adversarial interactions at police log inspections \cite{levy_data_2023}. Levy notes how, much like the cashiers in our study, drivers and law enforcement used the ELDs as a common enemy to turn these adversarial interactions into alliant ones \cite{levy_data_2023}. As these forms of relational patchworking emerge in other frontline contexts as well, it is of particular interest to CSCW scholars to understand the conditions that lead to relational changes and necessitate relational patchworking.

\section{Conclusion}

CSCW and organizational scholars have long studied how purportedly autonomous systems can create new forms of work, which are often invisibilized. Through interviews with frontline retail workers, we show how pseudo-autonomous systems --- which purport to automate tasks, but instead offset labor onto an unpaid labor source --- can similarly involve new forms of \textit{relational} work from employees. Faced with the responsibilities of monitoring customers, enforcing rules, and solving problems, workers become positioned as adversaries to customers, while having to still provide positive customer service. Relational patchworking techniques like scapegoating and hyperservicing allow workers to straddle the line between adversarialism and customer service. We turn to CSCW researchers to further explore the specific workplace conditions and technologies that necessitate or minimize these forms of relational patchworking in other workplaces, and how relationships between frontline workers and their customers or clients evolve as a result.


\section{Acknowledgements}
    We extend thanks to the retail workers who participated in our study for sharing their experiences and insights. We also thank Coworker.org and the moderators of the /r/retailhell Subreddit for their support in recruiting participants. We are grateful for comments and feedback from members of the Cornell AI, Policy, and Practice working group; from participants of the 2024 SEA/SAW Spring Meeting; and from José A. Guridi. This work was graciously supported by the John D. and Catherine T. MacArthur Foundation and the National Science Foundation Graduate Student Fellowship Program under Grant No. DGE–2139899.

\bibliographystyle{ACM-Reference-Format}
\bibliography{references}
\end{document}